\documentclass[%
 reprint,twocolumn,
superscriptaddress,
groupedaddress,
showpacs,preprintnumbers,
 amsmath,amssymb,amsthm
 aps
]{revtex4-1}
\usepackage{graphicx}
\usepackage{dcolumn}
\usepackage{bm}

\usepackage{xcolor}
\usepackage{latexsym}
\usepackage{filecontents}

\usepackage{natbib}

\begin{document}


\title{Sufficient condition for nonexistence of symmetric extension of qudits using Bell inequalities}
\author{Meenu Kumari$^{1,2}$, Shohini Ghose$^{1,3}$, and Robert B. Mann$^{1,2}$}
\affiliation{$^1$Institute for Quantum Computing, University of Waterloo, Canada N2L 3G1} 
\affiliation{$^2$Department of Physics and Astronomy, University of Waterloo, Canada N2L 3G1}
\affiliation{$^3$Department of Physics and Computer Science, Wilfrid Laurier University, Waterloo,
Canada N2L 3C5} 

\date{\today}

\begin{abstract}
We analyze the connection between Bell inequality violations and symmetric extendibility of quantum states. We prove that 2-qubit reduced states of multiqubit symmetric pure states do not violate the Bell Clauser-Horne-Shimony-Holt (CHSH) inequality. We then prove the more general converse that any 2-qubit state that violates the CHSH inequality cannot have a symmetric extension. We extend our analysis to qudits and provide a test for symmetric extendibility of 2-qudit states. We show that if a 2-qudit Bell inequality is monogamous, then any 2-qudit state that violates this inequality does not have a symmetric extension. For the specific case of 2-qutrit states, we use numerical evidence to conjecture that the Collins-Gisin-Linden-Massar-Popescu (CGLMP) inequality is monogamous.  Hence, the violation of the CGLMP inequality by any 2-qutrit state could be a sufficient condition for the non-existence of its symmetric extension.
 
\end{abstract}


\maketitle

\section{\label{sec:level1}Introduction}

Any 2-qudit quantum state $\rho_{AB}$ is said to have a symmetric extension if there exists a 3-qudit state $\rho_{ABB'}$ such that tracing over the qudit B or B$^\prime$ yields the same quantum state, that is, $\rho_{AB} = \rho_{AB'}$ \cite{Chen2014}. Symmetric extendibility of quantum states has been used in various areas of quantum information and quantum communication, such as detection of entanglement, determining entanglement distillability, and characterizing anti-degradable channels, to name a few \cite{Miguel2009,Fernando2012,myhr2009spectrum,Khatri2016}. It is therefore crucial to determine which states have a symmetric extension and which do not. Although semidefinite programming (SDP) \cite{vandenberghe1996semidefinite,doherty2005detecting} can do this numerically, it is a computationally expensive task. Thus it is desirable to have analytical  necessary and/or sufficient conditions to determine the symmetric extendibility of quantum states.  While necessary and sufficient conditions for the existence  of symmetric extensions have been obtained for 2-qubit states \cite{Chen2014},  finding the corresponding conditions for 2-qudit states remains  an open question (though a specific class of qudit states has been studied to this end \cite{Ranade2009PRA,Ranade2009JPhys,Johnson2013}).

In this paper, we provide a sufficient condition for the non-existence of symmetric extension for 2-qudit states based on 2-qudit Bell inequalities. Specifically, we establish the connection between Bell inequality violations and symmetric extendibility for qudit states by exploiting  the monogamy of Bell inequalities (some earlier work has used the existence of symmetric extension of quantum states to construct local hidden variable theories for these states, see \cite{Doherty2002,*Terhal2003,*Doherty2004}). We first focus on 2-qubit states and prove that 2-qubit reduced density matrices derived from multiqubit symmetric pure states can never violate the Bell Clauser-Horne-Shimony-Holt (CHSH) inequality (we will henceforth refer to this as the CHSH inequality). 
Next we prove the more general converse, namely that any 2-qubit state violating the CHSH inequality cannot have a symmetric extension. The result follows from the monogamy of the CHSH inequality \cite{Scarani2001,toner2009monogamy}. We generalize our proof to 2-qudit states to show that if a Bell inequality for 2-qudit states is monogamous, then its violation by a 2-qudit state implies that there cannot exist a 3-qudit  symmetric extension of the state. This is a sufficient condition for the non-existence of symmetric extension of a 2-qudit quantum state. 

Our results highlight the importance of monogamy in Bell inequalities. We thus explore the monogamous nature of the Collins-Gisin-Linden-Massar-Popescu (CGLMP) inequality, which is a Bell inequality for qudit states \cite{Collins2002}. We provide numerical evidence for the monogamy of the CGLMP inequality for 2-qutrit states. We conjecture that it is monogamous and thus conclude that the violation of the CGLMP inequality by a 2-qutrit state would imply that it does not have a  3-qutrit symmetric extension.

The paper is organized as follows. In section \ref{sec:level2}, we briefly discuss the CHSH and CGLMP Bell inequalities for qubits and qudits respectively, as well as symmetric extensions of quantum states. In section \ref{sec:level3}, we prove that 2-qubit reduced density matrices derived from multiqubit symmetric pure states can never violate the CHSH inequality. In section \ref{sec:level4}, we prove that any 2-qubit state that violates the CHSH inequality cannot have a symmetric extension. We then show that this proof can be simply extended to derive a sufficient condition for the symmetric non-extendibility of 2-qudit states. In section \ref{sec:level5}, we explore the CGLMP inequality and provide numerical evidence for the monogamous nature of the CGLMP inequality for  qutrits. From this, we conjecture that the CGLMP inequality is monogamous, and thus can be used to test the symmetric extendibility of 2-qutrit states.

\section{\label{sec:level2}Background}

\subsection{\label{sec:level2A}Bell inequalities}

Here, we introduce Bell inequalities for different quantum states. The CHSH and CGLMP inequalities are 2-qubit and 2-qudit Bell inequalities respectively. 
\medskip
 
\textbf{2-qubit states :} The CHSH correlation function, $\mathcal{B}(\rho)$, for any 2-qubit state, $\rho$ is 
\begin{equation}
\mathcal{B}(\rho) = \max_{A,B,A',B'} \langle AB+AB'+A'B-A'B' \rangle ,
\end{equation}
where $A$ and $A'$ are operators acting on the first qubit, and $B$ and $B'$ are operators acting on the second qubit. All four operators are such that their eigenvalues are $\pm 1$. 

Given any $\rho$, calculating $\mathcal{B}(\rho)$   is clearly an optimization problem. In \cite{Horodecki1995}, an analytical formula for the CHSH correlation function has been derived which does not involve any optimization. 
Consider a matrix $T$ whose elements are  
\begin{equation}
T_{ij}= \mbox{tr}(\rho \sigma_i \otimes \sigma_j).
\label{L6}
\end{equation} 
Let $U=T^{T}T$. Then,
\begin{equation}
\mathcal{B}(\rho)=2 \sqrt{u+v},
\end{equation}
where $u$ and $v$ are the largest and second largest eigenvalues of $U$. If $\mathcal{B}(\rho) > 2$, i.e., $u+v > 1$, then the state $\rho$ is said to have nonlocal correlations.

\textbf{2-qudit states :} The CGLMP inequality \cite{Collins2002} is the generalization of the Bell inequality for higher dimensional systems. We present this inequality for qutrits, for which the relevant operators  each have three outcomes, denoted $0$, $1,$ and $2$. Let $A_1$ and $A_2$ be the operators acting on the first qutrit  and $B_1$ and $B_2$  operators acting on the second qutrit.  The CGLMP correlation function, $\mathcal{I}_3(\rho)$ for a 2-qutrit state $\rho$, is
\begin{eqnarray}
\mathcal{I}_3(\rho)& = & P(A_1=B_1) +P(B_1=A_2+1)+P(A_2=B_2)\nonumber \\ &&+P(B_2=A_1)-P(A_1=B_1-1)-P(B_1=A_2)\nonumber \\ &&-P(A_2=B_2-1)-P(B_2=A_1-1).
\label{CGLMP1}
\end{eqnarray}
A convenient choice for the $A_1,A_2,B_1$ and $B_2$ operators is as follows \cite{Collins2002,acin2002}.
Let $\phi_k(j), \varphi_l(j)$, $j\in \{0,1,2\}$, $k,l \in \{1,2\}$ be 12 angles. Let $U(\vec{\phi_k})$ and $U(\vec{\varphi_l})$ be $3\times 3$ unitary operators whose diagonal elements are $\exp{(-\mathrm{i}\phi_k(j))}$ and $\exp{(-\mathrm{i}\varphi_l(j))}$, and off-diagonal elements are zero. Let $U_{\text{FT}}$ and $U^{*}_{\text{FT}}$ be the respective 3-dimensional discrete Fourier transform and inverse. The operators $A_k, B_l$, with $k,l \in \{1,2 \}$, are defined as 
\begin{eqnarray}
A_k = & U_{\text{FT}}(\vec{\phi}_k) U(\vec{\phi_k}), & \hspace{6mm} k \in \{1,2\} \nonumber \\
B_l = & U^*_{\text{FT}}(\vec{\varphi}_l)U(\vec{\varphi_l}), & \hspace{6mm} l \in \{1,2\}
\label{Operators}
\end{eqnarray} 
and their application is followed by a measurement in the $\{|0 \rangle, |1 \rangle, |2 \rangle \}$ basis. 
Thus,
\begin{eqnarray}
P(A_m=j,B_n=k)&=&\text{tr}(\Pi_j \otimes \Pi_k A_m \otimes B_n \rho A_m^{\dagger} \otimes B_n^{\dagger} ) \nonumber \\
\Rightarrow P(A_m=B_n)& = & \sum_{j=0}^2 P(A_m=j,B_n=j),
\label{Probabilities}
\end{eqnarray}
where $\rho$ is the 2-qutrit state. Using these probabilities in \eqref{CGLMP1}, we get the value of $\mathcal{I}_3$ as illustrated in Appendix A. The only variables here are the 12 angles, $\vec{\phi_1},\vec{\phi_2},\vec{\varphi_1}$ and $\vec{\varphi_2}$. Maximizing $\mathcal{I}_3$ over these angles 
yields the CGLMP inequality
\begin{equation}
\mathcal{I}_3 \leq  \underset{\vec{\phi}_k,\vec{\varphi}_l}{\text{max}}(\mathcal{I}_3)  \equiv \mathcal{B}_{CGLMP}(\rho) .
\label{CGLMP2}
\end{equation} 

For any local hidden variable (LHV) model, $\mathcal{I}_3(\rho) \leq 2$.  In contrast, for 
$\vert{\Psi}\rangle = \frac{1}{\sqrt{2+\gamma^2}}(|00 \rangle + \gamma |11 \rangle + |22 \rangle)$
the value of $\mathcal{B}_{CGLMP}(\vert\Psi \rangle)$   is $(1+\sqrt{11/3}) \approx  2.9149$ where $\gamma = 0.7923$ \cite{acin2002}. This is the maximal value of $\mathcal{I}_3$. 
\subsection{\label{sec:level2B}Symmetric extension of quantum states}

The condition 
\begin{equation}\label{symext}
\mbox{tr}(\rho_B^2) \geq \mbox{tr}(\rho_{AB}^2)-4\sqrt{\det(\rho_{AB})} ,
\end{equation} 
is necessary and sufficient for a 2-qubit state $\rho_{AB}$ to possess a symmetric extension, where $\text{tr}_A(\rho_{AB}) = \rho_B$ \cite{Chen2014}.  We are interested here in exploring the analogous situation for a 2-qudit state $\rho_{AB}$, which is said to have symmetric extension if there exists a  3-qudit state $\rho_{ABB'}$ such that
\begin{equation}
\mbox{tr}_{B} \rho_{ABB'} = \mbox{tr}_{B'} \rho_{ABB'} = \rho_{AB}. 
\end{equation}

\section{\label{sec:level3}Nonlocality of 2-qubit reduced states of multiqubit symmetric pure states}

Consider a state 
\begin{equation}
|\psi \rangle = \sum_{m=-j}^j c_m |j,m \rangle ,
\label{L3}
\end{equation}
where $|j,m \rangle$ are the eigenstates of the angular momentum operators $J^2$ and $J_z$ and $j,m$ are the angular momentum quantum numbers. The state $|\psi \rangle$ lies in a $2j+1$ dimensional  Hilbert space. This state belongs to the symmetric subspace of $2j$-qubit states (i.e., a symmetric combination of $N=2j$ spin-1/2 qubits). Multiqubit symmetric states are of special importance in quantum information; examples include the W state and the GHZ state.  We consider here $j \geq 3/2$, that is, $N \geq 3$ multiqubit symmetric  states. 

We denote by $\rho_{AA}$ an arbitrary 2-qubit state that is symmetric under pair exchange. We denote by $\varrho_{AA}$  the 2-qubit symmetric state (under pair exchange)  derived from the multiqubit symmetric pure state in Eq. \eqref{L3}. We obtain $\varrho_{AA}$ by tracing out any $(N-2)$  qubits from that multiqubit symmetric pure state. Furthermore, $|\psi \rangle$ can be seen as a symmetric purification of $\varrho_{AA}$.

We can obtain the 3-qubit symmetric extension of
$\varrho_{AA}$ by tracing out $(N-3)$ qubits from $|\psi \rangle$.  It will therefore satisfy the symmetric extendibility criterion
\begin{equation}
\mbox{tr}(\varrho_A^2) \geq \mbox{tr}(\varrho_{AA}^2)-4\sqrt{\det(\varrho_{AA})} ,
\label{L1} 
\end{equation}
and since  $\mbox{rank}(\varrho_{AA}) \leq 3$ we have
\begin{equation*}
\det(\varrho_{AA})=0.
\end{equation*}
Consequently Eq. \eqref{L1} becomes
\begin{equation}
\mbox{tr}(\varrho_A^2) \geq \mbox{tr}(\varrho_{AA}^2).
\label{L2} 
\end{equation}
It is crucial to note here that all $\rho_{AA}$'s are not guaranteed to possess such a symmetric extension.  

We briefly recapitulate the properties of $\varrho_{AA}$  \cite{Wang2002}. Any $\rho_{AA}$ takes the following form 
\begin{equation}
\begin{array}{r} \rho_{AA} =  \\ \end{array} \begin{bmatrix} v_+ & x_+^* & x_+^* & u^* \\ x_+ & w & y^* & x_-^* \\ x_+ & y & w & x_-^* \\ u & x_- & x_- & v_- 
\end{bmatrix}
\label{L5}
\end{equation}
in the basis $\{|00\rangle,|01\rangle,|10\rangle,|11\rangle \}$.  
Now if $\rho_{AA}=\varrho_{AA}$ derived from $|\psi\rangle$ in Eq. \eqref{L3}, then the matrix components are written as 
\begin{eqnarray}
v_{\pm} &=& \frac{N^2-2N+4 \langle J_z^2 \rangle \pm \langle J_z \rangle (N-1)}{4N(N-1)}, \nonumber \\
x_{\pm} &=& \frac{(N-1)\langle J_+ \rangle \pm \langle [J_+,J_z]_+ \rangle}{2N(N-1)}, \nonumber \\
w &=& \frac{N^2-4\langle J_z^2 \rangle}{4N(N-1)}, \nonumber \\
y &=& \frac{2\langle J_x^2 + J_y^2 \rangle - N}{2N(N-1)} = \frac{N^2-4\langle J_z^2 \rangle}{4N(N-1)} = w,\nonumber \\
u &=& \frac{\langle J_+^2 \rangle}{N(N-1)}.
\label{L4}
\end{eqnarray}
We now demonstrate that $\varrho_{AA}$ does not violate the CHSH inequality.
\\ \\
\textbf{Theorem 1:} $\mathcal{B}(\varrho_{AA}) \leq 2$, where matrix elements of $\varrho_{AA}$ are defined in Eqs. \eqref{L5} and \eqref{L4}, for $j\geq 3/2$.

\textbf{Proof :} For the 2-qubit state in Eq. \eqref{L5}, the $T$ matrix defined in Eq. \eqref{L6} is  
\begin{equation}
\begin{array}{r} T =  \\ \end{array} \begin{bmatrix} 2(w+\mbox{Re}(u)) & 2\mbox{Im}(u) & 2\mbox{Re}(x_+ - x_-) \\ 2\mbox{Im}(u) & 2(w-\mbox{Re}(u)) & 2 \mbox{Im}(x_+-x_-) \\ 2 \mbox{Re}(x_+ - x_-) & 2 \mbox{Im}(x_+-x_-) & 1-4w
\end{bmatrix}
\label{L7}
\end{equation}
It is clear that  $T$ is a symmetric matrix, and it is straightforward to show that its eigenvalues $\lambda_1$, $\lambda_2$ and $\lambda_3$ are real. Sorting them in order such that
\begin{equation}
\lambda_1^2 \leq \lambda_2^2 \leq \lambda_3^2  ,
\label{L8}
\end{equation} 
we find that  
\begin{equation}
\mathcal{B}(\varrho_{AA})=2\sqrt{\lambda_2^2 + \lambda_3^2}.
\label{L9}
\end{equation}
Furthermore $T$ in Eq. \eqref{L7} has unit trace and so 
\begin{equation}
 \lambda_1 + \lambda_2 + \lambda_3 = 1.
\label{L10}
\end{equation}
Squaring both sides of this equation yields, after some simplification
\begin{eqnarray}
\lambda_1^2 + \lambda_2^2 + \lambda_3^2  =
1- 2(\lambda_1 \lambda_2 + \lambda_2 \lambda_3 +  \lambda_1 \lambda_3) .
\label{L11}
\end{eqnarray}
From the properties of $3 \times 3$ matrices,
\begin{eqnarray}
&& \lambda_1 \lambda_2 + \lambda_2 \lambda_3 + \lambda_1 \lambda_3 = \mbox{Sum of all 2} \times \mbox{2 principal minors}\nonumber \\ 
&& = \begin{vmatrix} T_{11} & T_{12} \\ T_{21} & T_{22} \end{vmatrix} + \begin{vmatrix} T_{22} & T_{23} \\ T_{32} & T_{33} \end{vmatrix} + \begin{vmatrix} T_{11} & T_{13} \\ T_{31} & T_{33} \end{vmatrix}.
\label{L13}
\end{eqnarray}
Since $T$ is a symmetric matrix for $\varrho_{AA}$, Eq. \eqref{L13} becomes
\begin{eqnarray}
\lambda_1 \lambda_2 + \lambda_2 \lambda_3 + \lambda_1 \lambda_3 &=& T_{11}T_{22}+T_{22}T_{33}+T_{11}T_{33}\nonumber \\ && -T_{12}^2-T_{23}^2-T_{13}^2.
\label{L14}
\end{eqnarray}
Substituting matrix elements of $T$ from Eq. \eqref{L7} in Eq. \eqref{L14}, we get
\begin{equation}
\lambda_1 \lambda_2 + \lambda_2 \lambda_3 + \lambda_1 \lambda_3 = 4(w-3w^2-|u|^2-|x_+-x_-|^2).
\label{L15}
\end{equation}
Using Eq. \eqref{L2}, we have
\begin{eqnarray}
\mbox{tr}(\varrho_A^2)- \mbox{tr}(\varrho_{AA}^2) \geq 0.
\label{L16}
\end{eqnarray}
Now,
\begin{eqnarray}
\mbox{tr}(\varrho_{AA})&=& 1=v_++v_-+2w ,\nonumber \\
\varrho_A &=& \begin{bmatrix} v_++w & x_+^*+x_-^* \\ x_++x_- & v_-+w \end{bmatrix},\nonumber \\
\mbox{tr}(\varrho_{AA}^2) &=& v_+^2+v_-^2+2|u|^2+4(|x_+|^2+|x_-|^2+w^2),\nonumber \\
\mbox{tr}(\varrho_A^2) &=& (v_++w)^2+(v_-+w)^2+2|x_++x_-|^2  .
\label{L17}
\end{eqnarray}
Using Eq. \eqref{L17} in Eq. \eqref{L16}, we get
\begin{eqnarray}
w(v_++v_-)-|x_+-x_-|^2-w^2-|u|^2 \geq 0.
\label{L18}
\end{eqnarray}
Equation \eqref{L18} implies
\begin{eqnarray}
-w^2-|u|^2-|x_+-x_-|^2 &\geq & -w(v_++v_-)\nonumber \\
\Rightarrow w-3w^2-|u|^2-|x_+-x_-|^2 &\geq & w-2w^2\nonumber \\
& & -w(v_++v_-)\nonumber \\
& = & w(1-v_+-v_-) - 2w^2 \nonumber \\
& = & w\times2w - 2w^2  = 0\nonumber \\ & & \mbox{  (using tr(}\varrho_{AA})=1\mbox{)}\nonumber \\
\Rightarrow w-3w^2-|u|^2-|x_+-x_-|^2 &\geq & 0.
\label{L19}
\end{eqnarray}
Using Eq. \eqref{L19} in Eq. \eqref{L15}, we get
\begin{equation}
\lambda_1 \lambda_2 + \lambda_2 \lambda_3 + \lambda_1 \lambda_3 \geq 0.
\label{L20}
\end{equation}
Using Eq. \eqref{L20} in Eq. \eqref{L11}, we get
 \begin{eqnarray}
\lambda_1^2 + \lambda_2^2 + \lambda_3^2  =
1- 2(\lambda_1 \lambda_2 + \lambda_2 \lambda_3 +  \lambda_1 \lambda_3) \leq 1.
\label{L21}
\end{eqnarray}
Using Eq. \eqref{L21} in Eq. \eqref{L9} proves the result, namely $\mathcal{B}(\varrho_{AA})\leq 2$.

A recent paper showed that the violation of certain multipartite Bell-type inequalities (having terms containing only one and two body correlators) was indicative of multipartite entanglement \cite{Tura1256}. A particular class of permutation symmetric states was shown to exhibit maximum violation of these inequalities.  All two-body reduced states of these symmetric states were local in the considered scenario, and so the proposed generalized Bell inequalities are also not violated by 2-qubit reduced states of multi-qubit permutation symmetric states, consistent with our claim for the CHSH inequality.

\section{\label{sec:level4}No Symmetric extension of 2-qudit nonlocal states}

We build on the result of the previous section to explore the nonlocality of arbitrary 2-qubit states. We shall prove a more general result which holds for any 2-qubit state. Recall that CHSH correlation functions have been proven to be monogamous \cite{Scarani2001,toner2009monogamy}.  Specifically, if
$\rho_{ABC}$ is any three qubit state such that $\rho_{AB}$, $\rho_{BC}$ and $\rho_{AC}$ are its three 2-qubit reduced density matrices, then at most only one of these can violate the CHSH inequality. For example,
\begin{equation}
\mathcal{B}(\rho_{AB}) > 2 \Rightarrow \mathcal{B}(\rho_{BC}) \leq 2 \mbox{ and } \mathcal{B}(\rho_{AC}) \leq 2.
\label{L21a}
\end{equation}
Using this monogamy relation we prove the following theorem.  
\\

\textbf{Theorem 2 :} Any 2-qubit state that violates the CHSH inequality cannot possess a symmetric extension. \\ 

\textbf{Proof :} We will prove the theorem by contradiction. Let $\rho_{AB}$ be any two-qubit state for which $\mathcal{B}(\rho_{AB}) > 2$.

Suppose there exists a symmetric extension of $\rho_{AB}$, which is $\rho_{ABC}$. Then either of the following holds true :
\begin{eqnarray}
\hspace{2mm} 
\rho_{BC}  \equiv \mbox{tr}_{A}(\rho_{ABC}) = \rho_{AB} & \mbox{  or  } & 
\rho_{AC} \equiv \mbox{tr}_{B}(\rho_{ABC}) = \rho_{AB}, \nonumber \\
\end{eqnarray}
and so if $\rho_{AB}$ violates the CHSH inequality, either  $\rho_{BC}$ or $\rho_{AC}$ will 
also violate it, in contradiction with the monogamy relation of Eq. \eqref{L21a}. \hfill $\square$ \hfill \\

We specifically discuss the symmetric extendibility of pure states here because entangled pure states such as maximally entangled Bell states are the most useful resource in quantum computation. All entangled 2-qubit pure states are also nonlocal, which, using theorem 2,  implies that they cannot be symmetrically extended . We also see this using the criteria in Eq. \eqref{symext}. $\mbox{tr}(\rho_{AB}^2) = 1$ for pure states and $\mbox{det}{\rho_{AB}}=0$ (since pure states have rank 1).  Furthermore, the 1-qubit reduced density matrix (RDM) of any entangled  2-qubit density matrix must be mixed, and so $\mbox{tr}(\rho_{A}^2) < 1$. Hence
 \begin{eqnarray}\label{pure1}
\mbox{tr}(\rho_{A}^2) - \mbox{tr}(\rho_{AB}^2) + 4\sqrt{\mbox{det}\rho_{AB}} = \mbox{tr}(\rho_{A}^2) -1 < 0,
\end{eqnarray}
in contradiction with Eq. \eqref{symext}. Consequently no two-qubit entangled states satisfy the symmetric extendibility criterion. 

Theorem 2 can be generalized to establish a sufficient condition for non-extendibility of 2-qudit states. We prove the following theorem:\\

\textbf{Theorem 3 :} If a 2-qudit Bell inequality is monogamous, then any 2-qudit state that violates this inequality cannot have a symmetric extension. \\ 

\textbf{Proof :} We prove Theorem 3 by the method of contradiction along the lines of Theorem 2. Consider a 2-qudit  Bell inequality $\mathcal{B}(\rho) \leq N$. The upper bound $N$ will depend on the dimension of the qudits. Suppose that this 2-qudit Bell inequality is monogamous. This means that if  $\rho_{ABC}$ is any 3-qudit state such that $\rho_{AB}$, $\rho_{BC}$ and $\rho_{AC}$ are its three 2-qudit reduced density matrices, then at most only one of these can violate the 2-qudit Bell inequality. For example,
\begin{equation}
\mathcal{B}(\rho_{AB}) > N \Rightarrow \mathcal{B}(\rho_{BC}) \leq N \mbox{ and } \mathcal{B}(\rho_{AC}) \leq N.
\label{L21aa}
\end{equation}
Now suppose there exists a symmetric extension of the 2-qudit state $\rho_{AB}$, which is $\rho_{ABC}$. Then either of the following holds true :
\begin{eqnarray}
\hspace{2mm} 
\rho_{BC}  \equiv \mbox{tr}_{A}(\rho_{ABC}) = \rho_{AB} & \mbox{  or  } & 
\rho_{AC} \equiv \mbox{tr}_{B}(\rho_{ABC}) = \rho_{AB}, \nonumber \\
\end{eqnarray}
and so if $\rho_{AB}$ violates the 2-qudit Bell inequality then either of $\rho_{BC}$ or $\rho_{AC}$ will 
also violate it, in contradiction to the monogamy relation of Eq. \eqref{L21aa}. \hfill $\square$ \hfill \\

Thus, we have proved a sufficient condition for the non-existence of symmetric extension of 2-qudit states.

\section{\label{sec:level5}Symmetric extension of qutrit states}
We now apply the criterion provided in Theorem 3 to the case of  2-qutrit states. According to Theorem 3, we must first identify a monogamous 2-qutrit Bell inequality in order to test for 2-qutrit symmetric extendibility. To this end, we perform numerical studies 
of the monogamous nature of the CGLMP inequality Eq. \eqref{CGLMP1} for qutrit states (introduced in Sec. \ref{sec:level2A}). Based on our studies, we conjecture that the CGLMP inequality is monogamous. Given this conjecture, Theorem 3 implies that a violation of the CGLMP inequality by any 2-qutrit state is a sufficient condition for the non-existence of its 3-qutrit symmetric extension. 

\begin{figure}
\centering{\includegraphics[scale=0.44]{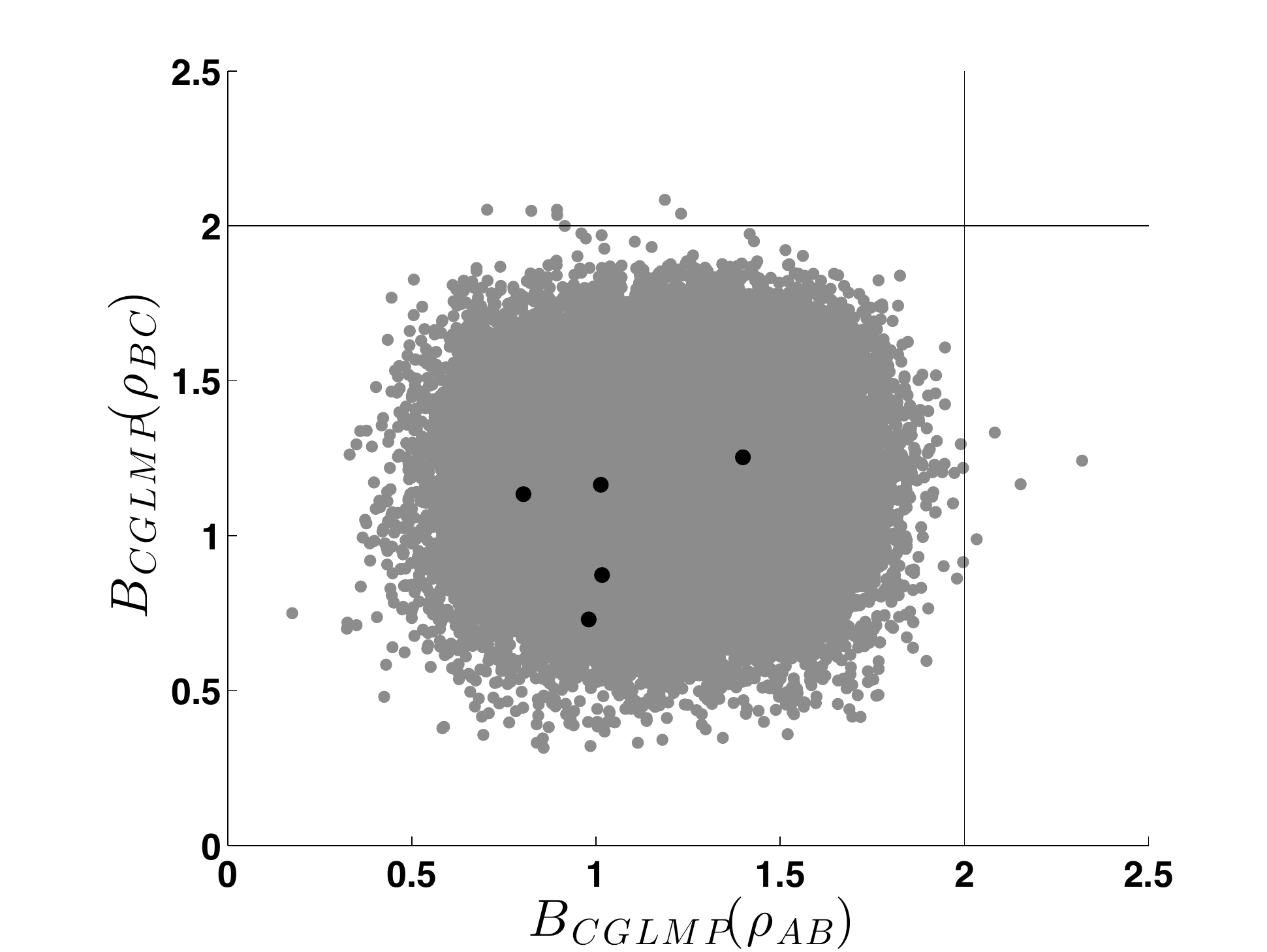}}
\caption{Maximum value of the CGLMP correlation function for 2-qutrit RDMs $\rho_{AB}$ and $\rho_{BC}$ of 50 000 random 3-qutrit pure states. States with $\mathcal{B}_{\text{CGLMP}}(\rho_{AC}) > 2$ are shown in black and states with $\mathcal{B}_{\text{CGLMP}}(\rho_{AC}) \leq 2$ in grey.}
\label{fig4}
\end{figure}

\begin{figure}
\centering{\includegraphics[scale=0.45]{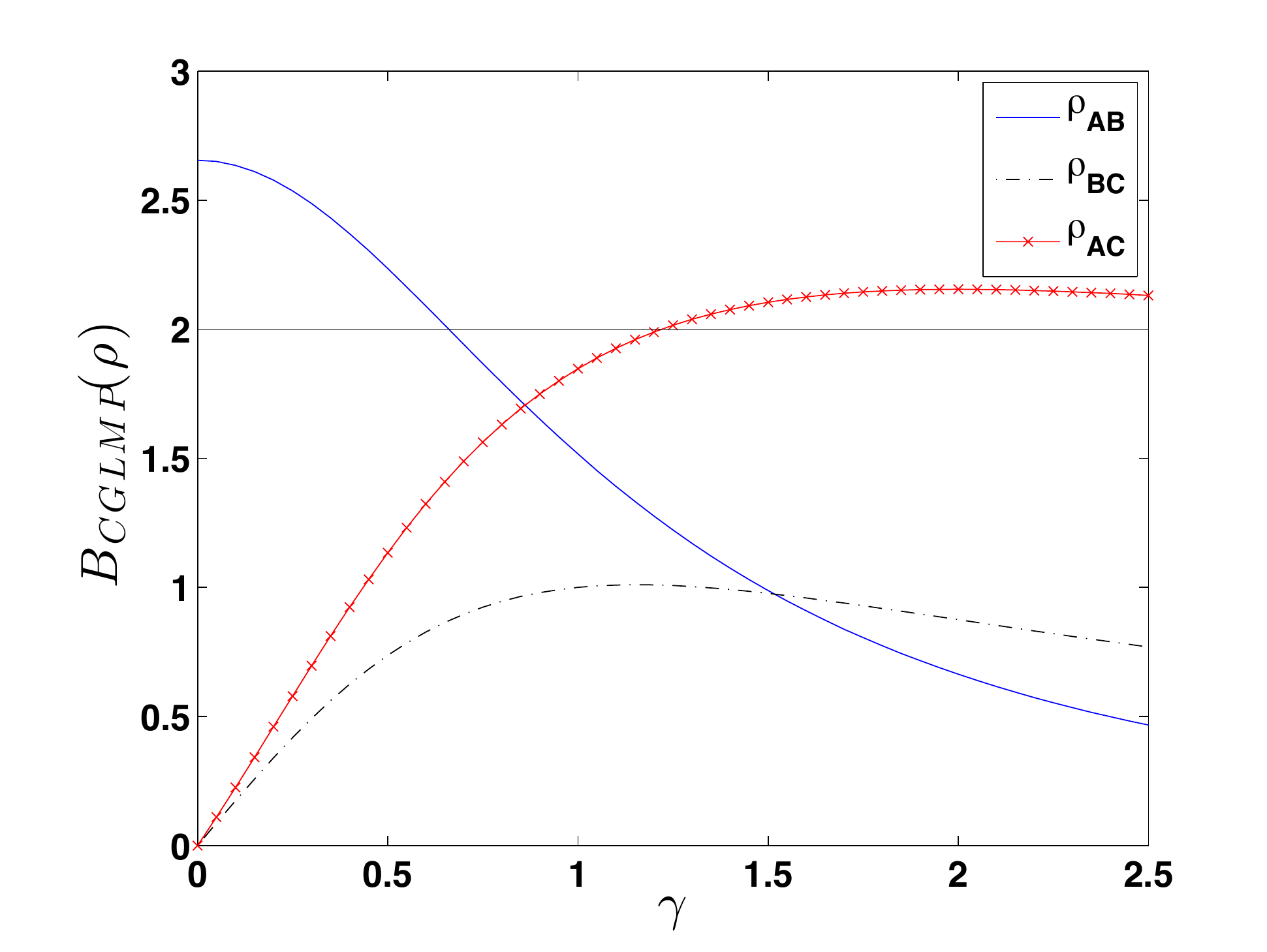}}
\caption{Maximum value of the CGLMP correlation function for the 2-qutrit RDMs of the 3-qutrit state in Eq. \eqref{3qutrit} as a function of
$\gamma$.}
\label{fig2}
\end{figure}

\begin{figure}
\centering{\includegraphics[scale=0.45]{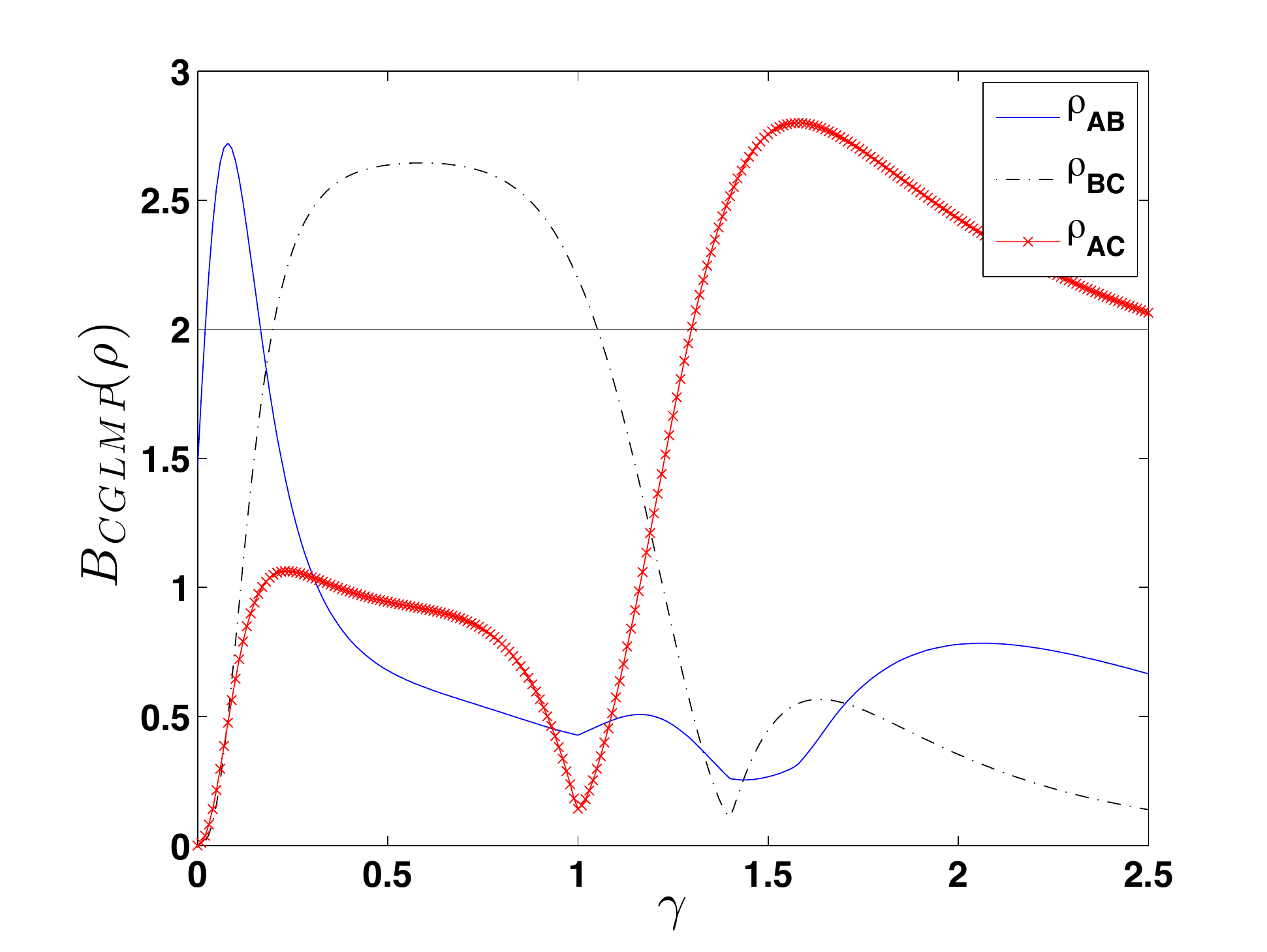}}
\caption{Maximum value of the CGLMP correlation function for the 2-qutrit RDMs of the 3-qutrit state in Eq. \eqref{3qutritb} as a function of
$\gamma$. }
\label{fig3}
\end{figure}

We performed a numerical search for the monogamy relation of the CGLMP inequality, analogous to Eq. \eqref{L21aa}, over 3-qutrit random pure states, $|\psi_{ABC}\rangle$. We used the method in \cite{marsaglia1972choosing} to uniformly sample 3-qutrit random pure states. Figure \ref{fig4} shows the CGLMP correlation function value for the three 2-qutrit reduced density matrices of 50 000 random 3-qutrit pure states, $|\psi_{ABC} \rangle$. The $X$ and $Y$ axes are the values of $\mathcal{B}_{\text{CGLMP}}(\rho_{AB})$ and $\mathcal{B}_{\text{CGLMP}}(\rho_{BC})$ respectively. Any 2-qutrit state $\rho$ is nonlocal if it violates the CGLMP inequality  $\mathcal{B}_{\text{CGLMP}}(\rho) \leq 2$. The 3-qutrit states with $\mathcal{B}_{\text{CGLMP}}(\rho_{AC}) > 2$ are represented by black dots and those for which 
$\mathcal{B}_{\text{CGLMP}}(\rho_{AC}) \leq 2$ with grey dots. As shown in the plot, there are no 3-qutrit states for which more than one 2-qutrit RDM violates the CGLMP inequality.
In order to test the conjecture further, we specifically construct 3-qutrit quantum states whose 2-qutrit RDMs show violation of the CGLMP inequality. We find that at most one 2-qutrit RDM violates the inequality for any 3-qutrit quantum state, thus respecting the monogamy relation given in Eq. \eqref{L21aa}. Here, we present the calculations for two such 3-qutrit states parametrized by $\gamma$ :
\begin{subequations}
\begin{equation}
\begin{split}
|\psi_1 \rangle = &\dfrac{1}{\sqrt{8+6\gamma^2}} \bigl( |000\rangle + |001\rangle + |002\rangle + |110\rangle + |111\rangle \\ & + |112 \rangle + |221\rangle + |222\rangle + \gamma \bigl( |010\rangle + |020\rangle + |112\rangle \\ & + |101\rangle + |121\rangle + |212\rangle \bigr) \bigr),
\end{split}
\label{3qutrit}
\end{equation}
\begin{equation}
\begin{split}
|\psi_2 \rangle = & \dfrac{1}{\sqrt{3+c_1^2+c_2^2+c_3^2}} \bigl( |000\rangle + |111\rangle + |222\rangle \\ & + c_1(|001\rangle + |002\rangle + |110\rangle + |112 \rangle + |220\rangle + |221\rangle ) \\ & + c_2 ( |100\rangle + |200\rangle + |011\rangle + |211\rangle + |022\rangle + |122\rangle )  \\ & + c_3( |010\rangle  + |020\rangle +  |101\rangle + |121\rangle + |202\rangle + |212\rangle ) \bigr),
\end{split}
\label{3qutritb}
\end{equation}
\end{subequations}
where $c_1=(10\gamma+0.01)^{-1}$,
$c_2 = -3\gamma \left( \gamma-1.4 \right) e^{-\gamma}$,
$c_3 = \gamma \left(\gamma-1 \right)$, and
$\rho_{ABC} =  |\psi \rangle \langle \psi | $. 
In Figs. \ref{fig2} and \ref{fig3}, we plot  
$\mathcal{B}_{\text{CGLMP}}$ for each of the three 2-qutrit RDMs of the states in Eq. \eqref{3qutrit} and Eq. \eqref{3qutritb} respectively, $\rho_{AB},\rho_{BC}$ and $\rho_{AC}$ as a function of the parameter $\gamma$. We see from  Fig. \ref{fig2} and Fig. \ref{fig3} that only one of the three 2-qutrit RDMs has $\mathcal{B}_{\text{CGLMP}}(\rho) > 2$ for any value of $\gamma$.
Based on our numerical studies, we make the following conjecture.
The CGLMP inequality for 2-qutrit states is  monogamous, that is, if $\rho_{ABC}$ is any 3-qutrit state such that $\rho_{AB}$, $\rho_{BC}$ and $\rho_{AC}$ are  its three 2-qutrit RDMs, at most one of these  violates the CGLMP inequality. For example,
\begin{eqnarray}
\mathcal{B}_{\text{CGLMP}}(\rho_{AB}) > 2 & \nonumber \\ \Rightarrow \mathcal{B}_{\text{CGLMP}}(\rho_{BC}) \leq 2 & \mbox{ and } \mathcal{B}_{\text{CGLMP}}(\rho_{AC}) \leq 2 
\label{L26}
\end{eqnarray}
with the same result holding for any permutation of $(A,B,C)$.
The above conjecture implies that any 2-qutrit state, $\rho_{AB}$, that violates the CGLMP inequality  does not possess a 3-qutrit symmetric extension. This follows from a simple application of Theorem 3. 

\section{\label{sec:level6}Discussion}
Theorem 2 shows that violation of the CHSH inequality is a sufficient condition for the non-existence of a symmetric extension of any 2-qubit state. This is a simple and practical method to test for the symmetric non-extendibility of 2-qubit states. A necessary and sufficient condition for the existence of symmetric extension of 2-qubit states has been previously given \cite{Chen2014}, with
this being specific only to 2-qubit states;  counterexamples demonstrate this does not hold for higher dimensional states  \cite{myhr2009spectrum}.

The analogous situation for qudit states has remained an open question. Here we have provided a test for determining when a 2-qudit state will not have a symmetric extension. Our criterion highlights the importance of monogamy of nonlocality.  We have found numerical evidence that the 2-qutrit CGLMP inequality is monogamous; in turn this provides an explicit method to test for the non-existence of symmetric extension of 2-qutrit states. Extensions to qudit states of higher dimensions could be obtained if higher-dimensional monogamous Bell inequalities can be identified.  Our work shows that nonlocality and symmetric extendibility are intrinsically linked, and provides motivation for future studies of monogamy of nonlocality. 
\section*{Acknowledgements}
MK and SG would like to thank Norbert L$\ddot{\text{u}}$tkenhaus for helpful discussions. This work was supported in part by the Natural Sciences and Engineering Research Council of Canada.
\subsection*{Appendix A}
The CGLMP correlation function, $\mathcal{I}_3(\rho)$ for a 2-qutrit state  is given in \eqref{CGLMP1}. Using \ref{Operators} and \ref{Probabilities}, $\mathcal{I}_3(\rho)$ can be written in the expanded form as (where addition in the index $j$ is modulo 2 addition) :
\begin{equation}
\begin{split}
\mathcal{I}_3(\rho) = \sum_{j=0}^2\text{tr}(\Pi_j \otimes \Pi_j A_1 \otimes B_1 \rho A_1^{\dagger} \otimes B_1^{\dagger} )  \\ 
+ \sum_{j=0}^2\text{tr}(\Pi_j \otimes \Pi_{j+1} A_2 \otimes B_1 \rho A_2^{\dagger} \otimes B_1^{\dagger} ) \\
+\sum_{j=0}^2\text{tr}(\Pi_j \otimes \Pi_j A_2 \otimes B_2 \rho A_2^{\dagger} \otimes B_2^{\dagger} ) \\
+ \sum_{j=0}^2\text{tr}(\Pi_j \otimes \Pi_j A_1 \otimes B_2 \rho A_1^{\dagger} \otimes B_2^{\dagger} ) \\
- \sum_{j=0}^2\text{tr}(\Pi_{j} \otimes \Pi_{j+1} A_1 \otimes B_1 \rho A_1^{\dagger} \otimes B_1^{\dagger} ) \\
-\sum_{j=0}^2\text{tr}(\Pi_j \otimes \Pi_{j} A_2 \otimes B_1 \rho A_2^{\dagger} \otimes B_1^{\dagger} ) \\
- \sum_{j=0}^2\text{tr}(\Pi_{j} \otimes \Pi_{j+1} A_2 \otimes B_2 \rho A_2^{\dagger} \otimes B_2^{\dagger} ) \\
- \sum_{j=0}^2\text{tr}(\Pi_{j+1} \otimes \Pi_{j} A_1 \otimes B_2 \rho A_1^{\dagger} \otimes B_2^{\dagger} ) .
\end{split}
\label{Expansion1}
\end{equation}

\bibliography{refer}

\end{document}